\documentclass[preprintnumbers, prd, twocolumn, showpacs, floatfix, preprintnumbers, letterpaper, superscriptaddress,nofootinbib]{revtex4}
\usepackage{graphicx,epsfig}
\usepackage{amsmath}
\usepackage {amssymb}
\usepackage{bm}
\usepackage{amsfonts}
\usepackage{epstopdf}
\usepackage{subfigure}
\usepackage{multirow}

\usepackage{relsize}
\usepackage{relsize}

\newcommand{\be}{\begin{equation}}
\newcommand{\ee}{\end{equation}}
\newcommand{\bea}{\begin{eqnarray}}
\newcommand{\eea}{\end{eqnarray}}

\begin{document}

\title{Towards  realistic $f(T)$ models with nonminimal torsion-matter coupling extension }

\author{Chao-jun Feng}
\email{fengcj@shnu.edu.cn}
\author{Fei-fei Ge}
\author{Xin-zhou Li}
\email{kychz@shnu.edu.cn}
\author{Rui-hui Lin}
\author{Xiang-hua Zhai}
\email{zhaixh@shnu.edu.cn}

\affiliation{Shanghai United Center for Astrophysics(SUCA), Shanghai
Normal University, 100 Guilin Road, Shanghai 200234, China
}%

\begin{abstract}
Using the observation data of SNeIa, CMB and BAO, we establish two concrete $f(T)$ models with nonminimal torsion-matter coupling extension.
We study in detail the cosmological implication of our models and find they are successful in describing the observation of the Universe, its large scale structure and evolution. In other words, these models
do not change the successful aspects of $\Lambda$CDM scenario under the error band of fitting values
as describing the evolution history of the Universe including radiation-dominated era,
matter-dominated era and the present accelerating expansion. Meanwhile, the significant advantage of these models is that they could avoid the cosmological constant problem of $\Lambda$CDM. A joint analysis is performed by using the data of CMB+BAO+JLA, which leads to $\Omega_{m0}=0.255\pm 0.010, \Omega_{b0}h^2=0.0221\pm 0.0003$ and $H_0=68.54\pm 1.27$ for model I and $\Omega_{m0}=0.306\pm 0.010, \Omega_{b0}h^2=0.0225\pm 0.0003$ and $H_0=60.97\pm 0.44$ for model II at 1$\sigma$ confidence level. The evolution of the decelaration parameter $q(a)$ and the effective equation of state $w_{DE}(a)$ are displayed. Furthermore, The resulted
age of the Universe from our models is consistent with the ages of the oldest globular
clusters. As for the fate of the Universe, model I results in a de Sitter accelerating
phase while model II appears a power-law one, even though $w_{DE0}< -1$ makes model I look like a phantom at present time.
\end{abstract}

\pacs{04.50.Kd, 98.80.-k, 95.36.+x}

\maketitle

\section{Introduction}\label{Introduction}

Einstein constructed the "Teleparallel Equivalence of General Relativity"
(TEGR) which is equivalent to General Relativity (GR) from the Einstein-Hilbert
 action\cite{Einstein,Unzicker,Moller,Pellegrini,Aldrovandi}.
 In TEGR, the curvatureless Weitzenb\"{o}ck connection takes the place of
 torsionless Levi-Civita one, and the vierbein is used as the fundamental
 field instead of the metric. In the Lagrangian of TEGR, the torsion scalar $T$
 by contractions of the torsion tensor takes the place of curvature scalar $R$.
 On the other hand, the extension of Einstein-Hilbert action with geometry-matter
 coupling $f(R,\mathcal{L}_M)$ was studied in Refs.\cite{Bertolami,Bertolami2,Harko1,Harko2}.
 An alternative modified model is the $f(R,\mathcal{T})$ theory \cite{Harko3}, where $\mathcal{T}$ is
 the trace of the matter energy-momentum tensor $\mathcal{T}_{\mu\nu}$. Following these lines, if one desires
 to modify gravity in TEGR, the simplest scheme is $f(T)$ theory \cite{Ferraro,Linder},
 whose important advantage is that the field equations are second order but not fourth order as in $f(R)$ theory.
 Furthermore, Harko and his colleagues \cite{Harko4} constructed an extension of $f(T)$
 gravity with a nonminimal torsion-matter coupling in the action, whose cosmological implication is rich and varied.
 Since $f(T)$ theories are known to violate local Lorentz invariance\cite{Barrow,Sotiriou}, particular choices of tetrad
 are important to get viable models in $f(T)$ cosmology, as has been noticed in Ref. \cite{Tamanini}.

According to Penrose's cosmic censorship hypothesis\cite{Penrose}, naked singularities should be abhorred. So a realistic and physically reasonable cosmological model should avoid the future singularities as far as possible\cite{Xi}, as well as accurately describe the history of the Universe. That is, besides accurately describing the expansion of the Universe at late time, a realistic model must be consistent with all observed facts such as proton-neutron ratio, baryon-photon ratio, the abundances of the light elements, the role of baryon perturbations, the evolution of the intergalactic gas and the galaxy formation. In order to explain these phenomena, cosmologists introduced the cosmological constant back again, and constructed a new standard model--$\Lambda$CDM, also known as the concordance model, which is intended to satisfy all the main observations such as type Ia supernovae (SNeIa), cosmic microwave background radiation (CMB), the large scale structure (LSS) and some observations at early time. Although the cosmological constant accounts for almost 70\% the whole energy density at present in $\Lambda$CDM, the value is still too small to be explained by any current fundamental theories. Lacking underlying theoretical foundations, the particular value of cosmological constant is just selected phenomenologically, which means the model is highly sensitive to the value of model parameter, and this results in the so-called fine-tuning problem. This problem is considered as the biggest issue for almost all cosmological models. In order to alleviate this troublesome problem, various dynamical dark energy theories have been proposed and developed these years, such as quintessence \cite{Peebles,Li} and phantom \cite{Caldwell,Li2}, in which the energy composition is dependent on time. But these exotic fields are still phenomenological, lacking theoretical foundations. Besides adding unknown fields, there is another kind of theories known as modified gravity, which uses alternative gravity theory instead of Einstein theory, such as $f(R)$ theory \cite{Nojiri,Du}, $f(T)$ theory \cite{Ferraro,Linder}, MOND cosmology \cite{Zhang}, Poincar\'{e} gauge theory \cite{Li3,Ao, Ao2}, and de Sitter gauge theory \cite{Ao3}. Among these theories, $f(T)$ theory with nonminimal torsion-matter coupling has a solid theoretical motivation, but to our knowledge there has not yet been any concrete model till now that can be compared with observations.

 In this paper, using the observation data of SNeIa, CMB and BAO,
 we establish a class of detailed models, which can not only explain the issue of the accelerating  expansion of the Universe, but also be consist with the evolution history of radiation-dominated era
 and matter-dominated era. As for the fate of the Universe, it is de Sitter evolving in model I whereas it is power-law evolving in model II. There are no future singularities in both of our models.

 The paper is organized as follows; in section I we introduce the motivation of the paper. In section II, we shall give a brief review of $f(T)$ theory with nonminimal torsion-matter coupling extension and two concrete models. The latest observation constraints and fitting results for our models are given in section III. In section IV, we shall study the cosmological implications for our models. Finally, section V is devoted to the conclusion and discussion.

 \section{Theory and Models}
\subsection{ A Brief Review of the Theory}
We can find a set of smooth basis vector fields $\hat{e}_{(\mu)}$ in different patches of the manifold $\mathcal{M}$ and make sure things are well-behaved on the overlaps as usual, where greek indices run over the coordinate of spacetime. The set of vectors $\textbf{e}_A$ comprising an orthonormal basis is known as tetrad or vierbein, where latin indices run over the tangent space $T_p$ at each point $p$ in $\mathcal{M}$. Any vector can be expressed as  linear combinations of basis vector, so we have

\begin{equation}
\hat{e}_{(A)}=e_A ^{\hspace{0.2cm}\mu}\hat{e}_{(\mu)}
\end{equation}
where the components $e_A^{\hspace{0.2cm}\mu}$ form a $4\times 4$ invertible matrix. We will also refer to $e_A^{\hspace{0.2cm}\mu}$ as the vierbein in  accordance with usual practice of blurring the distinction between objects and their components. The vectors $\hat{e}_{(\mu)}$ in terms of  $\hat{e}_{(A)}$ are
\begin{equation}
\hat{e}_{(\mu)}=e^A _{\hspace{0.2cm}\mu}\hat{e}_{(A)}
\end{equation}
where the inverse vierbeins $e^A _{\hspace{0.2cm}\mu}$ satisfy
\begin{equation}
e^A_{\hspace{0.2cm}\mu}e_B^{\hspace{0.2cm\mu}}=\delta_B^A,e_A^{\hspace{0.2cm}\mu}e^A_{\hspace{0.2cm}\nu}=\delta_\nu^\mu.
\end{equation}
Therefore, the metric is obtained from $e^A_{\hspace{0.2cm}\mu}$
\begin{equation}
g_{\mu\nu}=\eta_{AB}e^A_{\hspace{0.2cm}\mu}e^B_{\hspace{0.2cm}\nu},
\end{equation}
or equivalently
\begin{equation}\eta_{AB}=g_{\mu\nu}e_A^{\hspace{0.2cm}\mu}e_B^{\hspace{0.2cm}\nu},
\end{equation}
and the root of the metric determinant is given by $|e|=\sqrt{-g}=\det(e^A_{\hspace{0.2cm}\mu})$.

In TEGR, one uses the standard Weitzenb\"{o}ck's connection defined as
\begin{equation}
\Gamma^\alpha_{\mu\nu}=e_A^{\hspace{0.2cm}\alpha}\partial_\nu e^A_{\hspace{0.2cm}\mu}=-e^A_{\hspace{0.2cm}\mu}\partial_\nu e_A^{\hspace{0.2cm}\alpha}.
\end{equation}
And the covariant derivative $\mathrm{D}_\mu$ satisfies the equation
\begin{equation}
\mathrm{D}_\mu e^A_{\hspace{0.2cm}\nu}=\partial_\mu e^A_{\hspace{0.2cm}\nu}-\Gamma^\alpha_{\nu\mu}e^A_{\hspace{0.2cm}\alpha}=0.
\end{equation}
Then the components of the torsion and contorsion tensors are given by
\begin{eqnarray}
T^\alpha_{\hspace{0.2cm}\mu\nu}&=&\Gamma^\alpha_{\nu\mu}-\Gamma^\alpha_{\mu\nu}=e_A^{\hspace{0.2cm}\alpha}(\partial _\mu e^A_{\hspace{0.2cm}\nu}-\partial_\nu e^A_{\hspace{0.2cm}\mu}),\\
K^{\mu\nu}_{\hspace{0.3cm}\alpha}&=&-\frac 1 2 (T^{\mu\nu}_{\hspace{0.3cm}\alpha}-T^{\nu\mu}_{\hspace{0.3cm}\alpha}-T_\alpha^{\hspace{0.2cm}\mu\nu}).
\end{eqnarray}
By introducing another tensor
\begin{equation}
S_\alpha^{\hspace{0.2cm}\mu\nu}=\frac 1 2(K^{\mu\nu}_{\hspace{0.3cm}\alpha}+\delta ^\mu _\alpha T^{\beta\nu}_{\hspace{0.3cm}\beta}-\delta ^\nu_\alpha T^{\beta\mu}_{\hspace{0.3cm}\beta}),
\end{equation}
we can define the torsion scalar as
\begin{equation}
T\equiv T^\alpha_{\hspace{0.2cm}\mu\nu}S_\alpha^{\hspace{0.2cm}\mu\nu}.
\end{equation}

Inspired by Harko et. al. who proposed a nonminimal torsion-matter coupling in the gravitational action\cite{Harko4},  to describe the evolution history of the Universe including the era before matter-radiation equality, we include the radiation in the action
\begin{equation}
S=\frac 1 {16\pi G}\int|e|((1+f_1(T))T+(1+f_2(T))\mathcal{L}_M+\mathcal{L}_r)\mathrm{d}^4 x,
\end{equation}
where $\mathcal{L}_M$ and $\mathcal{L}_r$ are the Lagrangian densities of matter and radiation, respectively. More general consideration should include torsion-radiation coupling, that is, the third term should be $(1+f_3(T))\mathcal{L}_r$, but here $f_3=0$ has been taken for simplicity, and one will see in Sections III and IV that with such simplification our models can well explain the acceleration and the evolution history of the Universe. Applying the action principle with respect to the vierbein field, one can obtain the equation of motion as
\begin{eqnarray}
& &\frac 1 4 e_A^{\hspace{0.2cm}\rho}T(1+f_1)+(1+f_1+Tf_1^{'}+f_2^{'}\mathcal{L}_M)(|e|^{-1}\partial_\mu(|e|e_A^{\hspace{0.2cm}\alpha}S_\alpha^{\hspace{0.2cm}\rho\mu})\nonumber\\
&+&e_A^{\hspace{0.2cm}\alpha}T^\mu_{\hspace{0.2cm}\nu\alpha}S_\mu^{\hspace{0.2cm}\rho\nu})
+4(2f_1^{'}+Tf_1^{''}+f_2^{''}\mathcal{L}_M)(\partial_\mu T)e_A^{\hspace{0.2cm}\alpha}S_\alpha^{\hspace{0.2cm}\rho\mu}\nonumber\\
&+&4f_2^{'}e_A^{\hspace{0.2cm}\alpha}S_\alpha^{\hspace{0.2cm}\rho\mu}(\partial_\mu \mathcal{L}_M)
=4\pi G((1+f_2)\mathcal{T}^\rho_{\alpha(M)}+\mathcal{T}^\rho_{\alpha(r)})e_A^{\hspace{0.2cm}\alpha},\nonumber\\
\end{eqnarray}
where the prime indicates derivative with respect to torsion scalar $T$, and $\mathcal{T}^\rho_{\alpha(M)}$ and $\mathcal{T}^\rho_{\alpha(r)}$ are the matter and radiation energy-momentum tensor with
\begin{equation}
\frac{\delta(|e|\mathcal{L}_M)}{\delta e^A_{\hspace{0.2cm}\rho}}=-2|e|\mathcal{T}^\rho_{\alpha(M)}e_A^{\hspace{0.2cm}\alpha},\frac{\delta(|e|\mathcal{L}_r)}{\delta e^A_{\hspace{0.2cm}\rho}}=-2|e|\mathcal{T}^\rho_{\alpha(r)}e_A^{\hspace{0.2cm}\alpha},
\end{equation}
respectively.

For a flat Friedmann-Robertson-Walker metric in Cartesian coordinates,
\begin{equation}
ds^2=-dt^2+a(t)^2(dx^i)^2
\end{equation}
where $a(t)$ is the scale factor, the diagonal tetrad $e^A_{\hspace{0.2cm}\mu}=\mathrm{diag}(1,a,a,a)$ is a good choice to get viable models\cite{Tamanini}, with the local Lorentz noninvariance of $f(T)$ theories considered. And the torsion scalar $T=-6H^2$, where $H=\dot{a}/a$ is the Hubble parameter. The equation of motion then reads
\begin{widetext}
\begin{eqnarray}
H^2&=&\frac 1 3 (1+f_2)\rho_M+\frac 1 3\rho_r-2H^2f^{'}_2\mathcal{L}_M
-\frac 1 6(Tf_1+12H^2(f_1+Tf^{'}_1)),\\
\dot{H}&=&-\frac{\frac 1 2(1+f_2)(\rho_M+p_M)+\frac 1 2 (\rho_r+p_r)+f^{'}_2H\mathcal{\dot{L}}_M}{1+f_1+Tf^{'}_1+f^{'}_2\mathcal{L}_M-12H^2(2f_1{'}+Tf^{''}_1)-12H^2f^{''}_2\mathcal{L}_M}\nonumber\\
\end{eqnarray}
\end{widetext}
where $\rho_M,p_M,\rho_r$ and $p_r$ are the densities and pressures of matter and radiation, and $\rho_M=\rho_m+\rho_b$ for realistic models consisting of cold dark and baryon matters. The overdot means derivative with respect to time. Here and after, we use the units $8\pi G=1$.

From Eqs. (16) and (17) one can consider the divergence of the matter and radiation energy-momentum tensors
\begin{eqnarray}
& &\frac 1 3 \mathrm{D}_\mu \mathcal{T}^{0\mu}_{(r)}+\frac 1 3 (1+f_2)\mathrm{D}_\mu \mathcal{T}^{0\mu}_{(M)}\nonumber\\
&=&\frac 1 3[\dot{\rho}_r+3H(\rho_r+p_r)]+\frac 1 3(1+f_2)[\dot{\rho}_M+3H(\rho_M+p_M)]\nonumber\\
&=&2H\dot{H}f^{'}_2(\mathcal{L}_M+2\rho_M).
\end{eqnarray}
Now if one considers matter as perfect fluid and chooses the matter Lagrangian density as $\mathcal{L}_M=-2\rho_M$ \cite{Harko4,Gron,Bertolami3}, matter and radiation conservation
\begin{equation}
\dot{\rho} _r+4H\rho_r=0, \dot{\rho }_M+3H\rho_M=0
\end{equation}
will automatically hold regardless of $f_1$ and $f_2$.

 \subsection{Two Concrete Models}
We consider two concrete models as follows:
\begin{equation}
  f_1=\begin{cases}\frac{12B H_0^4}{T^2}\\H_0(-T)^{-\frac 1 2}\end{cases},\quad f_2=\frac{-2A H_0^2}{\Omega_{M0}T}\quad \text{for} \begin{array}{c}\text{Model I}\\\text{Model II}\end{array},\nonumber
\end{equation}
where $A, B$ are dimensionless parameters, and $H_0$ and $\Omega_{M0}\equiv \frac{\rho_{M0}}{3H_0^2}$ are the current values of $H$ and the matter density parameter $\Omega_M$. In fact, for model II the contribution of $f_1$ to the equation of motion (16) and (17) vanishes. With $T=-6H^2$, our models can be equivalently expressed as $f_1=\frac B 3 \left (\frac {H_0}H\right )^4$ for model I and $f_2=\frac A {3\Omega_{M0}}\left (\frac {H_0} H\right )^2$ for both models. It is obvious that at current time, we have $f_1=\frac B 3$ and $f_2=\frac A {3\Omega_{M0}}$. In other words, the present values of $f_1$ and $f_2$ depend only on two dimensionless parameters $A$ and $B$ but not $H_0$. One will see in next section the best fitting values of $A=0.188$, $B=0.510$ for model I and $A=0.633$ for model II. Therefore, our models do not contain any parameter that needs fine tuning. Friedmann equation (16) can be unifiably rewritten as
\begin{equation}
E^2=\frac 1 2 (\Omega_{M0}a^{-3}+\Omega_{r0}a^{-4})\left [1+\sqrt{1+\frac{4(Aa^{-3}+B)}{(\Omega_{M0}a^{-3}+\Omega_{r0}a^{-4})^2}}\right ]
\end{equation}
with the constraint
\begin{equation}
\Omega_{M0}+\Omega_{r0}+A+B=1,
\end{equation}
where $E\equiv\frac H {H_0}$, $\Omega_{r0}\equiv \frac{\rho_{r0}}{3H_0^2}$, and $B\neq 0$ and $B=0$ correspond to model I and model II, respectively.

Let us introduce the effective dark energy density $\rho_{DE}$, then Eq. (20) becomes
\begin{equation}
E^2=\Omega_{M0}a^{-3}+\Omega_{r0}a^{-4}+\Omega_{DE}(a)
\end{equation}
where
\begin{eqnarray}
\Omega_{DE}(a)\equiv \frac{\rho_{DE}}{3H_0^2}
&=&\frac 1 2 \Big [\sqrt{(\Omega_{M0}a^{-3}+\Omega_{r0}a^{-4})^2+4(Aa^{-3}+B)}\nonumber\\
&-&(\Omega_{M0}a^{-3}+\Omega_{r0}a^{-4})\Big ]
\end{eqnarray}
and
\begin{equation}
\Omega_{DE0}=A+B=1-\Omega_{M0}-\Omega_{r0}.
\end{equation}
By using the conservation law, we get the equation of state as

\begin{eqnarray}
& &w_{DE}(a)=-1-\frac 1 3\frac{d\ln \rho_{DE}}{d\ln a}\nonumber\\
&=&-1-\frac{\Omega_{M0}+\frac 4 3\Omega_{r0}a^{-1}-\frac{\Omega_{M0}^2+\frac 7 3\Omega_{M0}\Omega_{r0}a^{-1}+\frac 4 3\Omega_{r0}^2a^{-2}+2Aa^3}{[(\Omega_{M0}+\Omega_{r0}a^{-1})^2+4(Aa^3+Ba^6)]^{\frac 1 2}}}
{[(\Omega_{M0}+\Omega_{r0}a^{-1})^2+4(Aa^3+Ba^6)]^{\frac 1 2}-(\Omega_{M0}+\Omega_{r0}a^{-1})}\nonumber\\
\end{eqnarray}
and its corresponding expression at present time
\begin{eqnarray}
w_{DE0}&=&-1\nonumber-\frac 1 {2\Omega_{DE0}}\\
&\times&\left(\Omega_{M0}+\frac 4 3\Omega_{r0}-\frac{\Omega_{M0}^2+\frac 7 3\Omega_{M0}\Omega_{r0}+\frac 4 3 \Omega_{r0}^2+2A}{1+\Omega_{DE0}}\right )\nonumber\\
\end{eqnarray}

By taking the fitting values of $\Omega_{m0}=0.255\pm 0.010, \Omega_{b0}h^2=0.0221\pm 0.0003, A=0.188\pm 0.048$ (see next section), we get $w_{DE0}=-1.019\pm 0.035$ for model I. And for model II, $w_{DE0}=-0.6124\pm 0.014$ with the fitting values of $\Omega_{m0}=0.306\pm 0.010,  \Omega_{b0}h^2=0.0225\pm 0.0003$. Furthermore, the evolutions  of $w_{DE}(a)$ are illustrated in Figs. 1 and 2 for models I and II, respectively.

\begin{figure}[h]
\subfigure{
\includegraphics[width=0.4\textwidth,angle=0]{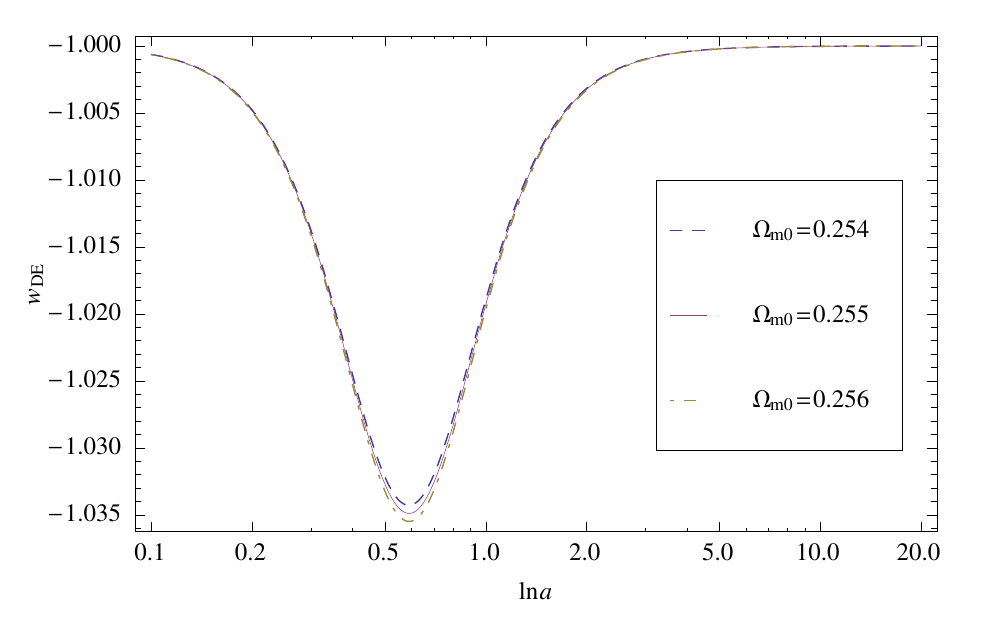}}
\subfigure{
\includegraphics[width=0.4\textwidth,angle=0]{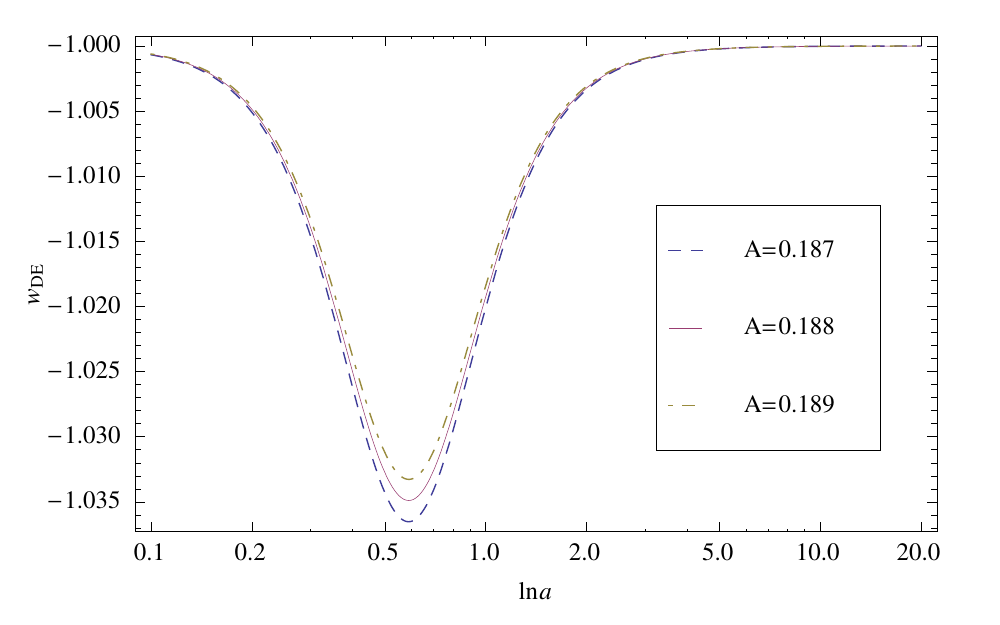}}
\caption{Evolution of equation of state with different parameter values for model I. In the upper $A$=0.188, and in the bottom $\Omega_{m0}=0.255$.}
\end{figure}

\begin{figure}[h]
\begin{center}
\includegraphics[width=0.4\textwidth,angle=0]{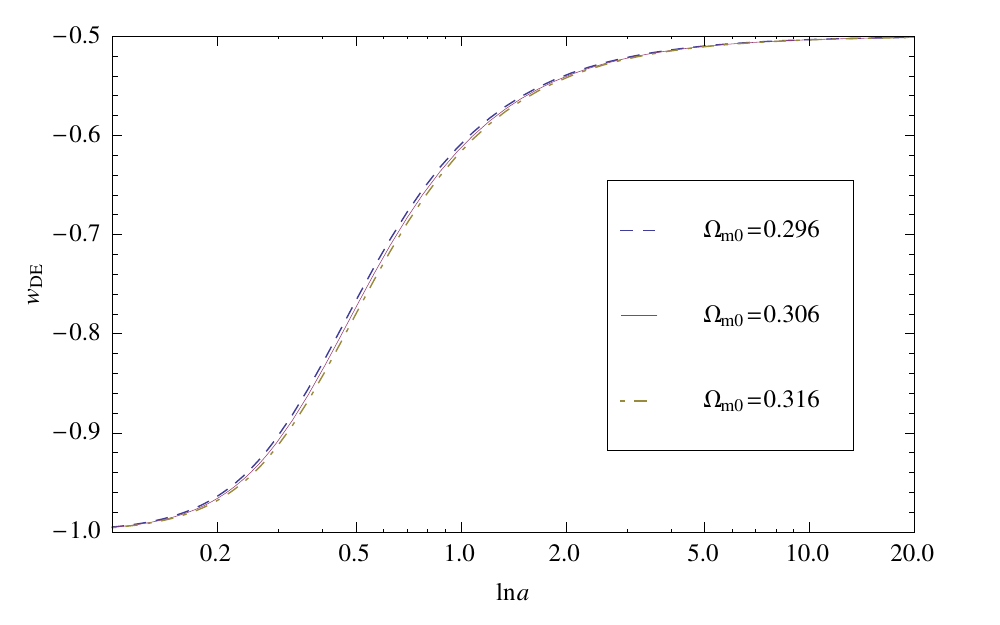}
\caption{Evolution of equation of state with different parameter values for model II. }
\end{center}
\end{figure}

\section{Latest Observation Constraints and Fitting Results}
\subsection{SNeIa Constraints}
In the following, we fit models I and II by using the ¡°joint light-curve analysis" (JLA)  sample, which contains $740$  spectroscopically confirmed type Ia supernovae with high quality light curves. The distance estimator in this analysis assumes hat supernovae with identical color, shape and galactic environment have on average the same intrinsic luminosity for all redshifts. This hypothesis is quantified by a linear model, yielding a standardized distance modulus\cite{Betoule:2014frx,Shafer:2015kda}
\begin{equation}\label{equ:modulusobs}
	\mu_{\text{obs}} = m_{\text{B}} - (M_{\text{B}} - \alpha \cdot s + \beta \cdot C + P \cdot \Delta_M)  \,,
\end{equation}
where $m_{\text{B}}$ is the observed peak magnitude in rest-frame B band, $M_{\text{B}}, s, C$ are the absolute magnitude, stretch and color measures, which are specific to the light-curve fitter employed, and $P(M_* >10^{10} M_\odot)$ is the probability that the supernova occurred in a high-stellar-mass host galaxy. The stretch, color, and host-mass coefficients ($\alpha, \beta, \Delta_M$, respectively) are nuisance parameters that should be constrained along with other cosmological parameters. On the other hand, the distance modulus predicted from a cosmological model  for a supernova at redshift $z$ is given by
\begin{equation}\label{equ:modulusmod}
	\mu_{\text{model}} (z, \vec\theta)= 5\log_{10} \left[ \frac{D_L(z)}{10\text{pc}}\right]\,,
\end{equation}
where $\vec\theta$ are the cosmological parameters in the model, and $D_L(z)$ is the luminosity distance
\begin{equation}\label{equ:lum}
	D_L(z) = (1+z)  \frac{c}{H_0} \int_0^z \frac{dz'}{E(z')} = (1+z)r_A(z)\,,
\end{equation}
for a flat FRW Universe. Here $r_A(z)$ is the comoving angular diameter distance. The $\chi^2$ statistic is then calculated in the usual way
\begin{equation}\label{equ:chi2sn}
	\chi^2_{\text{SN} } = (\vec \mu_{\text{obs}}  - \vec \mu_{\text{model}})^T \mathbf{C_{\text{SN}}}^{-1}  (\vec \mu_{\text{obs}}  - \vec \mu_{\text{model}})\,,
\end{equation}
with $\mathbf{C_{\text{SN}}}$ the covariance matrix of $\vec \mu_{\text{obs}}$.

\subsection{Cosmic Microwave Background  Data and Constraint}

The CMB temperature power spectrum is  sensitive to the matter density, and it also measures precisely the angular diameter distance at the last-scattering surface, which is defined as
\begin{equation}\label{equ:angu}
	\theta_{*} = \frac{r_s(z_*)}{r_A(z_*)} \,,
\end{equation}
where $r_s(z)$ is the comoving sound horizon
\begin{equation}
  r_s(z_*) = \frac{c}{H_0}\int_0^{a(z_*)} \frac{c_s(a)}{a^2 E(a)} da \,,
\end{equation}
with the sound speed $c_s(a)$ given by
\begin{equation}
  c_s(a) = \bigg[ 3 \left( 1+ \frac{3\Omega_{b0}h^2}{4\Omega_{\gamma 0}h^2} a \right) \bigg]^{-1/2} \,.
\end{equation}
Usually, $\theta_*$ is approximated based on the fitting function of $z_*$ given in Ref.~\cite{Hu:1995en}:
\begin{equation}
  z_* = 1048 \left[1+0.00124(\Omega_{b0}h^2)^{-0.738}\right]\left[ 1+ g_1(\Omega_{m0}h^2)^{g_2}\right] \,,
\end{equation}
 where
\begin{equation}
  g_1 = \frac{0.0783(\Omega_{b0}h^2)^{-0.238}}{1+39.5(\Omega_{b0}h^2)^{0.763}} \,, \quad
  g_2 = \frac{0.560}{1+21.1(\Omega_{b0}h^2)^{1.81} } \,.
\end{equation}
In CosmoMC package, the approximated $\theta_*$ is denoted as $\theta_{\text{MC}}$. In this paper, we fix $\Omega_{\gamma0} = 2.469\times 10^{-5} h^{-2}$,  and then the total radiation energy density, namely the sum of photons and relativistic neutrinos is  given by $\Omega_{r0} = \Omega_{\gamma0}(1+0.2271 N_{\text{eff}}) $,
where $ N_{\text{eff}}$ is the effective number of neutrino species, and the current standard value is $ N_{\text{eff}}=3.046$.  In the following, we use the Planck measurement of the CMB temperature fluctuations and the WMAP measurement of the large-scale fluctuations of the CMB polarization. This CMB data are often denoted by "Planck + WP". The geometrical constraints inferred from this data set are the present value of baryon density $\Omega_{b0}h^2$ and dark matter $\Omega_{m0}h^2$, as well as $100\theta_{\text{MC}}$. Thus, the $\chi^2 $ of the CMB data is constructed as
\begin{equation}\label{equ:chi2cmb}
	\chi^2_{\text{CMB}} = (\nu - \nu_{\text{CMB}})^T \mathbf{C}^{-1}_{\text{CMB}} (\nu - \nu_{\text{CMB}})\,,
\end{equation}
where $\nu_{\text{CMB}}^T = (\Omega_{b0}h^2, \Omega_{m0}h^2, 100\theta_{\text{MC}}) ^T= (0.022065, 0.1199, 1.04131)^T$, and the best fit covariance matrix for $\nu$ is given by\cite{Betoule:2014frx,Ade:2013zuv}
\begin{eqnarray*}
\mathbf{C}_{\text{CMB}} & = & 10^{-7}\left(\begin{array}{ccc}
    0.79039 & -4.0042 & 0.80608\\
    -4.0042 & 66.950 & -6.9243\\
    0.80608 & -6.9243 & 3.9712\end{array}\right)\, .
\end{eqnarray*}
after marginalized over all other parameters.

\subsection{Baryon Acoustic Oscillations Data and Constraint}
The BAO measurement provides a standard ruler to probe the angular diameter distance versus redshift by performing a spherical average of their scale measurement, which contains the angular scale and the redshift separation: $d_z = r_s(z_d)/D_V(z) $, where $r_s(z_d)$ is the comoving sound horizon at the baryon drag epoch,  and $D_V(z)$ is given by
\begin{equation}
  D_V(z) \equiv \bigg[ r_A^2(z)\frac{ c z }{H(z)}\bigg]^{1/3} \,.
\end{equation}
The redshift of the drag epoch can be approximated by the following fitting formula,
\begin{equation}
  z_d = \frac{ 1291(\Omega_{m0}h^2)^{0.251} }{ 1+ 0.659(\Omega_{m0}h^2)^{0.828} } \bigg[ 1+ b_1(\Omega_{b0}h^2)^{b_2}\bigg] \,,
\end{equation}
with
\begin{eqnarray}
  b_1 &=& 0.313(\Omega_{m0}h^2)^{-0.419} \left[1 + 0.607(\Omega_{m0}h^2)^{0.674}\right], \nonumber\\
  b_2 &=& 0.238(\Omega_{m0}h^2)^{0.223}  \,.
\end{eqnarray}
see, Ref. \cite{Eisenstein:1997ik}.  In the following, we will use the measurement of the BAO scale from Ref.~\cite{Beutler:2011hx,Padmanabhan:2012hf,Anderson:2012sa} and then the $\chi^2 $ of the BAO data is constructed as
\begin{equation}\label{equ:chi2bao}
	\chi^2_{\text{BAO}} = (d_z - d_z^{\text{BAO}})^T\mathbf{C}^{-1}_{\text{BAO}} (d_z - d_z^{\text{BAO}}) \,,
\end{equation}
with $d_z^{\text{BAO}}=(d_{0.106},d_{0.35},d_{0.57})^T=(0.336,0.1126,0.07315)^T$, and the covariance matrix, also see Ref.~\cite{Betoule:2014frx}
\begin{eqnarray*}
C_{\text{BAO}}^{-1} & = & \left(\begin{array}{ccc}
    4444 & 0 &0 \\
    0 & 215156 &0 \\
    0&0& 721487\end{array}\right)\,.
\end{eqnarray*}

\subsection{Fitting Results}
In Table \ref{table:best}, we present the  best-fit parameters by using the data of CMB+BAO+JLA, and  also quote their $1-\sigma$ bounds from the approximate Fisher Information Matrix. Here the fitting results of $\Lambda$CDM is comparable with those in Ref.\cite{Shafer:2015kda} where $\Lambda$CDM is also used as a reference. One can see that the fitting results of Model I is closer to $\Lambda$CDM than those of Model II. Fig. 3 illustrates the constraints on $\Omega_{m0}$ and $A$ from 1$\sigma$ to $3\sigma$ confidence level for model I.

\begin{table}[h]
\centering
  \begin{tabular}{c|c|c|c}
  \hline
  \hline
  \multirow{2}{*}{ Parameters} & \multicolumn{2}{c}{ Cosmological Models  } \\
  \cline{2-4}
  & Model I &  Model II & $\Lambda CDM$  \\
  \hline
  \hline
  $\Omega_{m0}$
  & $0.255\pm 0.010$  	& $0.306\pm0.010$ &  $0.257\pm 0.009$ \\
    $A$
  & $0.188\pm 0.048$ 	& $-$  &  $-$			     \\
   \hline
   $H_0$
  & $68.54\pm 1.27$ & $60.97\pm0.44$ & $68.03\pm 0.74$ \\
  $\Omega_{b0}h^2$
  & $0.0221\pm 0.0003$ & $0.0225\pm0.0003$ &  $0.0221\pm 0.0002$ \\
  \hline
  $\alpha$
  & $0.141\pm 0.006$ & $0.139 \pm 0.006$& $0.141\pm 0.006$ \\
  $\beta$
  & $3.106\pm 0.075$ & $3.076\pm0.074$ & $3.102\pm 0.075$ \\
  $M_B$
  & $-19.10\pm 0.033$ & $-19.25\pm0.022$  &$-19.109\pm 0.026$ \\
  $\Delta M$
  & $-0.070\pm 0.023$ & $-0.076\pm0.023$ & $-0.070\pm 0.023$ \\
  \hline
  \hline
  $\chi^2_{min}/d.o.f$
  & $683.846/738$ & $729.429/739$  & $684.081/739$\\
  \hline
  \hline
  \end{tabular}
  \caption{\label{table:best} Best fitting parameters for the two models.}
\end{table}

\begin{figure}[h]
\begin{center}
\includegraphics[width=0.4\textwidth,angle=0]{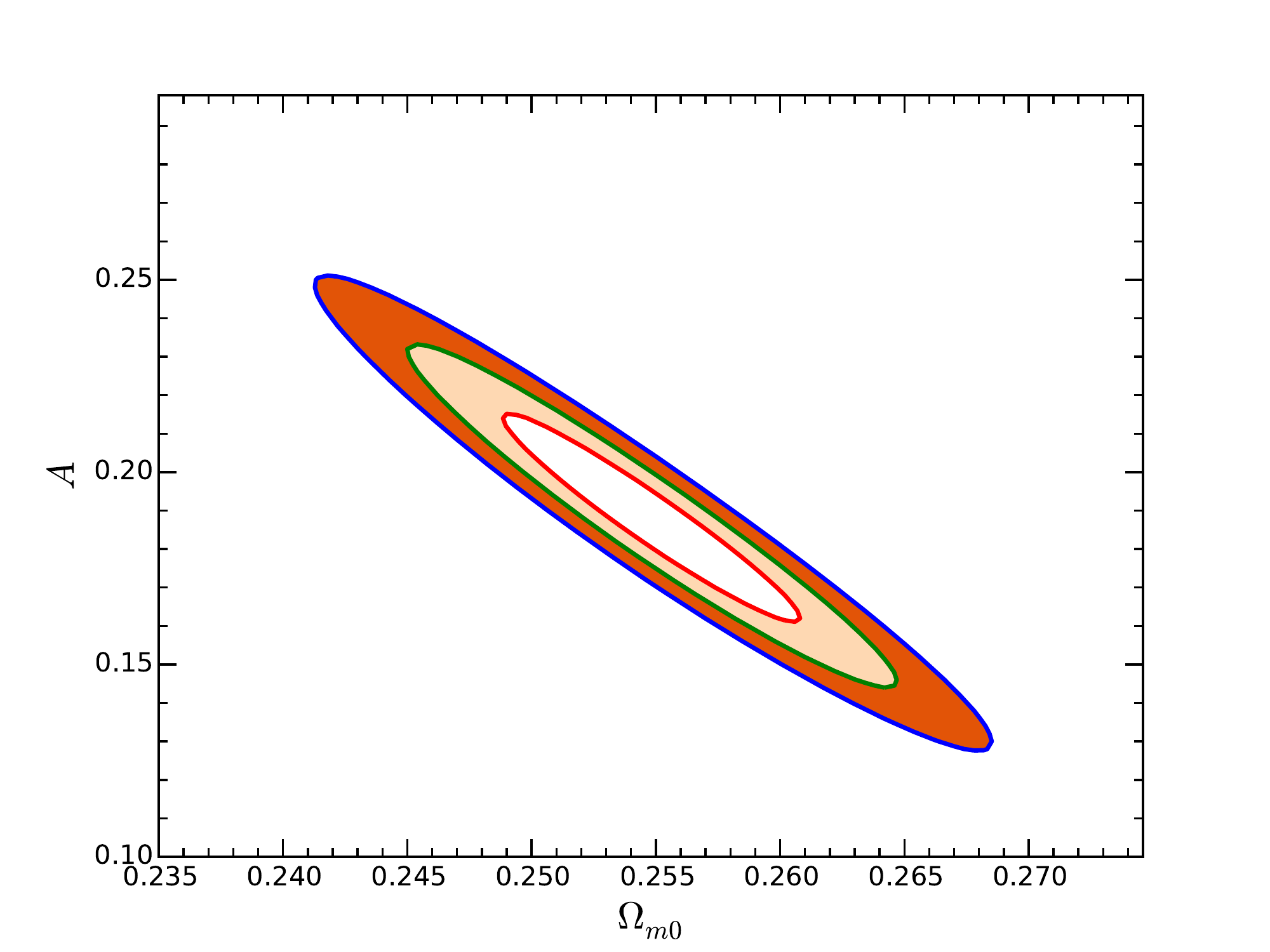}
\caption{Constraints on $\Omega_{m0}$ and $A$ from $1\sigma$ to $3\sigma$ confidence level obtained by using JLA SNe Ia+CMB+BAO for model I, while other parameters take their best fitting values.}
\end{center}
\end{figure}

 \section{Cosmological implication}
 \subsection{Evolution before matter-radiation equality Epoch}
 Let us retrospect the evolution history of standard cosmology from redshift of about $10^{12}$ to matter-radiation equality epoch. When redshift $z>10^{12}$, the baryon-antibaryon annihilate and their energy is transferred to the photon field. During this era, a tiny asymmetry takes place between baryon and antibaryon because of CP violation in $\mathrm{K}^\circ$ decay. At redshift $z\approx 10^{12}$, baryon-antibaryon pair generation from background radiation can be in progress and the Universe was glutted with these pairs. The opacity of the Universe for weak interactions become unity at $z\approx 10^{10}$, which lead up to a neutrino barrier, similar to the photon barrier at $z\approx 1000$.

 When redshift $z>10^9$, the $e^{\pm}$ annihilate and their energy is diverted into the photon field. At $z\approx10^9$, $e^{\pm}$ pair generation from background radiation can be underway and the Universe is congested with $e^{\pm}$ pairs. At redshift $z\approx 3\times 10^8$, the high energy photons in the tail of the Planck distribution are energetic enough to dissociate light nuclei such as $^2 \mathrm{D}$, $^3 \mathrm{T}$, $^3\mathrm{He}$, $^4\mathrm{He}$ and $^7\mathrm{Li}$. At $z\gg 4\times 10^4h^2$, the Universe is radiation-dominated. At $z_{eq}\approx 2.9\times 10^4h^2$, the matter density becomes equal to that of the radiation, which indicates the beginning of the current matter-dominated epoch and the start of structure formation.

 It is easy to proof that the thermal history of models I and II recover the usual one before the equipartition epoch. At the equipartition epoch, the scale factor
 \begin{equation}
 a_{eq}=\frac{\Omega_{\gamma 0}(1+0.2271N_{eff})}{\Omega_{M0}}.
 \end{equation}
Using the fitting values, we have
\begin{equation}
a_{eq}=\begin{cases}2.94\times10^{-4}\\2.42\times10^{-4} \end{cases}\quad \text{for} \begin{array}{c}\text{Model I}\\\text{Model II}\end{array}.
\end{equation}
Before the equipartition epoch, Friedmann equation is reduced to
\begin{equation}
E^2=(1+\delta(a))(\Omega_{M0}a^{-3}+\Omega_{r0}a^{-4})
\end{equation}
where $\delta(a)$ is a tiny minority,
\begin{equation}
\delta(a)\leq \delta_{eq}=\begin{cases}1.31\times10^{-11}\\1.67\times10^{-11} \end{cases}\quad \text{for} \begin{array}{c}\text{Model I}\\\text{Model II}\end{array}.
\end{equation}
Therefore, models I and II recover the usual evolution of the early Universe.

\subsection{The Evolution for redshift $6<z<z_{eq}$}
In redshifts $6<z<z_{eq}$, there are three critical epochs in the evolution as follows:

(i) The CMB temperature varies with redshift $z$ as $T_r=2.7255(1+z)\mathrm{K}$, so that $T_r\approx 4000\mathrm{K}$ at $z=1469$. The pregalactic gas was 50\% ionised at $z\approx1469$ and this epoch is known as recombination. When $z\lesssim 1469$, the universal plasma recombined to form neutral atoms. The abundance of light elements is fixed by primordial nucleosynthesis, which should be consistent with the observation.

(ii) At $z\geq 1000$, the Universe became opaque to Thomson scattering. If there exists no further scattering of the photons of the background radiation, $z=1000$ becomes the last scattering surface so that the fluctuations imprinted on the radiation at this epoch decide the spatial fluctuations in the intensity of CMB observed today.

(iii) The redshift of post-recombination epoch stretch from 1000 to zero. During this epoch, the structures such as galaxies and clusters of galaxies should be formed. The earlier phase of the post-recombination epoch, $6<z<1000$, is referred to as the dark ages. We have learned a considerable amount of the early development of the perturbations from CMB.

Due to its high nonlinearity, it is almost impossible to find the exact solutions of equation (20). Nevertheless, using series expansion and by analytic continuation, we can analytically describe the whole evolution history of the Universe.

From Eq.(20), the relation between $a(t)$ and $t$ is
\begin{equation}
\frac 2 3 a^{\frac 3 2}\left (\sum^{\infty}_{n=0}(-1)^n c_n a^{3n}\right )=\Omega^{\frac 1 2}_{M0}H_0 t
\end{equation}
during the matter-dominated era, and the coefficients of the expansion are
\begin{equation}
c_0=1,\quad c_n=\frac{A_n}{(n!)(2n+1)2^n \Omega_{M0}^{2n}},(n\geq 1)
\end{equation}
where
\begin{eqnarray}
A_1&=&A, \quad A_2=7A^2-4B\Omega_{M0}^2,\nonumber\\
A_3&=&99A^3-84AB\Omega_{M0}^2,\nonumber\\
A_4&=&2145A^4-2376A^2B\Omega_{M0}^2+336B^2\Omega_{M0}^4,\nonumber\\
A_5&=&62985A^5-85800A^3B\Omega_{M0}^2+23760AB^2\Omega_{M0}^4,\nonumber\\
&\cdots& \cdots
\end{eqnarray}
Similarly, the relation between $a(t)$ and $t$ for $\Lambda$CDM is
\begin{equation}
\frac 2 3 a^{\frac 3 2}\left (\sum^{\infty}_{n=0}(-1)^n \tilde{c}_n a^{3n}\right )=\Omega^{\frac 1 2}_{M0}H_0 t
\end{equation}
during the matter-dominated epoch, where the coefficients of the expansion are
\begin{equation}
\tilde{c}_0=1, \tilde{c}_n=\frac{(2n-1)!!}{n!(2n+1)2^n}\left (\frac{\Omega_{\Lambda 0}}{\Omega_{M 0}}\right )^n \quad (n\geq 1)
\end{equation}

It is easy to demonstrate that the convergent radius of the series in Eq. (45) are
\begin{equation}
r_{conv}=\begin{cases}\left ( \frac{A+\sqrt{A^2+B\Omega_{M0}^2}}{2B}\right )^{\frac 1 3}\\\left ( \frac {\Omega_{M0}^2} 4\right )^{\frac 1 3}\end{cases},\quad \text{for} \begin{array}{c}\text{Model I}\\ \text{Model II}\end{array}.
\end{equation}
Using the fitting values, we have
\begin{equation}
r_{conv}=\begin{cases}0.458\\0.375\end{cases},\quad \text{for} \begin{array}{c}\text{Model I}\\ \text{Model II}\end{array}.
\end{equation}

From Eqs. (45)-(49), we can obtain the time $t$ for corresponding $a$ in the convergent range of the series. To take $a=\frac 1 7$, we have
\begin{equation}
t_{z=6}=\begin{cases}\frac {2\Omega_{M0}^{-\frac 1 2}H_0^{-1}} 3 \sum_{n=0}^{\infty}(-1)^n c_n \left(\frac 1 7\right )^{3n+\frac 3 2}
\\\frac {2\Omega_{M0}^{-\frac 1 2}H_0^{-1}} 3 \sum_{n=0}^{\infty}(-1)^n \tilde{c}_n \left(\frac 1 7\right )^{3n+\frac 3 2}\end{cases}\quad \text{for} \begin{array}{c}\text{Models I,II}\\ \Lambda\text{CDM}\end{array}.
\end{equation}
Using the fitting values, we get
\begin{equation}
t_{z=6}=\begin{cases}0.0654H_0^{-1}\\0.0593H_0^{-1}\\0.0651H_0^{-1}\end{cases}\quad \text{for}\begin{array}{c}\text{Model I}\\ \text{Model II}\\ \Lambda\text{ CDM}\end{array}.
\end{equation}

The difference of scale factor $\Delta a$ at time $t$ between our models and $\Lambda$CDM can also be considered through
\begin{equation}
\sum_{n=0}^{\infty}(-1)^n\Delta_i c_n[a(1+\Delta a)]^{3n+\frac 3 2}=\sum_{n=0}^{\infty}(-1)^n \tilde{c}_n a^{3n+\frac 3 2}
\end{equation}
where $\Delta_i (i=1,2)$ are the ratios of $\Lambda$CDM to our models for $\Omega_{M0}^{\frac 12 }H_0$. Using the fitting values, we have
\begin{equation}
\Delta_i=\begin{cases}0.997\\1.017\end{cases}\quad   \text{for} \begin{array}{c}\text{Model I}\\\text{Model II}\end{array}.
\end{equation}
Between the matter-radiation equality epoch and the dark ages, $1470>z>6$, we can easily proof that
\begin{equation}
|\Delta a|<\left |\left (\frac 1 {\Delta_i}\right )^{\frac 2 3}-1\right |.
\end{equation}
Numerical calculation shows that
\begin{equation}
1.62\times 10^{-3}<\Delta a<1.68\times 10^{-3} \quad \text{for model I},
\end {equation}
and
\begin{equation}
-1.06\times 10^{-2}>\Delta a>-1.13\times10^{-2} \quad \text{for model II}.
\end{equation}

Therefore, our models do not change the $\Lambda$CDM scenario including primordial nucleosynthesis, CMB and dark ages under error band of fitting values. Especially, using the fitting values of $\Omega_{b0}h^2$, we have the baryon to photon ratio
\begin{equation}
\eta=\frac{N_b}{N_{\gamma}}=\begin{cases}6.13\times10^{-10}\\6.25\times10^{-10} \end{cases}\quad \text{for} \begin{array}{c}\text{Model I}\\\text{Model II}\end{array},
\end{equation}
where $N_b$ and $N_{\gamma}$ are number densities of baryons and photons, respectively. It is known that a cosmological model with $\eta \leq 10^{-10}$ or $\eta \geq 10^{-9}$ could not explain the primordial [D/H] ratio. The values of $\eta$ in our models are reasonable to explain the abundance of the light elements.  The fact that the number of photons is far more than that of baryons leads to ionize most of the neutral hydrogen in the intergalactic medium for $z>1470$.

\subsection{The Evolution for redshift $0\leq z\leq 6$}
As is known to all, the interval $0<z<6$ may be termed as the observable Universe of galaxies. The galaxy formation involves a large number of complex non-linear effects. To study the large-scale structure formation, we should combine our models with hydrodynamic codes which follow the dynamical and thermal history of the diffuse integalactic gas on principle. Fortunately, the $\Lambda$CDM model of large-scale structure formation has been considered by using the semi-analytic approach\cite{Longair}. Furthermore, we can still proof that the thermal history of our models is consistent with $\Lambda$CDM under error band of fitting values for observed data in the interval $0<z<6$.

For $z<1.18(1.66)$ for Model I(II), the series in Eq. (45) is divergent, so it cannot be used again. However, we can consider new expression by analytic continuation.
If we expand $a(t)$ at $a_{*}=a(t_{*})$ after the matter-dominated epoch, the relation between $a$ and $t$ is
\begin{equation}
\sum_{n=1}^{\infty} (-1)^n c_n(a_{*})(a-a_{*})^n=\Omega^{\frac 12 }_{M0}H_0(t-t_{*}).
\end{equation}
The coefficients $c_n(a_{*})$ of the expansion are
\begin{eqnarray}
c_1(a_{*})&=&-\frac{\sqrt{2a_{*}}}{\mu},\nonumber\\
c_n(a_*)&=&-\frac{(a_{*}^3A_n(a_{*})+\mu^2B_n(a_{*}))c_1(a_{*})}
{2(n!)a_{*}^{n-1}\mu^{2n-2}(\mu^2-1)^{2n-3}\Omega_{M0}^{2n-4+2[\frac n 2]}},(n\geq 2)\nonumber\\
\end{eqnarray}
where $[x]$ is the integer part of $x$, $\mu=(1+(1+(4Aa_{*}^3+4Ba_{*}^6)\Omega_{M0}^{-2})^{\frac 12})^{\frac 12 }$ and
\begin{widetext}
\begin{eqnarray}
A_2(a_{*})&=&-2A-8Ba_{*}^3,\quad B_2(a_{*})=\Omega_{M0}^2,\nonumber\\
A_3(a_{*})&=&2A\Omega_{M0}^2+(62A^2+2B\Omega_{M0}^2)a_{*}^3+232ABa_{*}^6+224B^2a_{*}^{9},\nonumber\\
B_3(a_{*})&=&\Omega_{M0}^4+24A\Omega_{M0}^2a_{*}^3-(10A^2-78B\Omega_{M0}^2)a_{*}^6+16ABa_{*}^9-64B^2a_{*}^{12},\nonumber\\
A_4(a_{*})&=&6A\Omega_{M0}^6-3(14A^2\Omega_{M0}^4-2B\Omega_{M0}^6)a_{*}^3+3(422A^3\Omega_{M0}^2-422AB\Omega_{M0}^4)a_{*}^6-3(120A^4-1264A^2B\Omega_{M0}^2-408B^2\Omega_{M0}^4)a_{*}^{9}\nonumber\\
&-&3(664A^3B-3512AB^2\Omega_{M0}^2)a_{*}^{12}+3(1248A^2B^2+2400B^3\Omega_{M0}^2)a_{*}^{15}+2304AB^3a_{*}^{18}-3072B^4a_{*}^{21},\nonumber\\
B_4(a_{*})&=&3\Omega_{M0}^8-18A\Omega_{M0}^6a_{*}^3+3(219A^2\Omega_{M0}^4-203B\Omega_{M0}^6)a_{*}^6+3(280A^3\Omega_{M0}^2+248AB\Omega_{M0}^4)a_{*}^9 \nonumber\\
&+&3(964A^2B\Omega_{M0}^2+1100B^2\Omega_{M0}^4)a_{*}^{12}+9312AB^2\Omega_{M0}^2a_{*}^{15}+4992B^3\Omega_{M0}^2a_{*}^{18},\nonumber\\
A_5(a_{*})&=&30A\Omega_{M0}^8+3(190A^2\Omega_{M0}^6+10B\Omega_{M0}^8)a_{*}^3-3(3804A^3\Omega_{M0}^4-3348AB\Omega_{M0}^6)a_{*}^6\nonumber\\
&+&3(11014A^4\Omega_{M0}^2-28380A^2B\Omega_{M0}^4+3158B^2\Omega_{M0}^6)a_{*}^{9}+3(23800A^5+24400A^3B\Omega_{M0}^2-110952AB^2\Omega_{M0}^4)a_{*}^{12}\nonumber\\
&+&3(138936A^4B-138960A^2B^2\Omega_{M0}^2-86376B^3\Omega_{M0}^4)a_{*}^{15}+3(238784A^3B^2-133184AB^3\Omega_{M0}^2)a_{*}^{18}\nonumber\\
&+&3(453376A^2B^3+24832B^4\Omega_{M0}^2)a_{*}^{21}+688128B^5a_{*}^{27}+1677312AB^4a_{*}^{24},\nonumber\\
B_5(a_{*})&=&15\Omega_{M0}^{10}+300A\Omega_{M0}^8a_{*}^3-3(1812A^2\Omega_{M0}^6-1584B\Omega_{M0}^8)a_{*}^6+3(3392A^3\Omega_{M0}^4-10624AB\Omega_{M0}^6)a_{*}^9\nonumber\\
&+&3(18130A^4\Omega_{M0}^2-9684A^2B\Omega_{M0}^4-41614B^2\Omega_{M0}^6)a_{*}^{12}-3(1560A^5-83280A^3B\Omega_{M0}^2+135672AB^2\Omega_{M0}^4)a_{*}^{15}\nonumber\\
&-&3(8344A^4B-58800A^2B^2\Omega_{M0}^2+33496B^3\Omega_{M0}^4)a_{*}^{18}-3(48256A^3B^2-215680AB^3\Omega_{M0}^2)a_{*}^{21}\nonumber\\
&+&3(32256A^2B^3+161280B^4\Omega_{M0}^2)a_{*}^{24}+172032AB^4a_{*}^{27}-49152B^5a_{*}^{30},\cdots \cdots
\end{eqnarray}
\end{widetext}

Similarly, the relation between $a(t)$ and $t$ for $\Lambda$CDM is
\begin{equation}
\sum_{n=1}^{\infty}\tilde{c}_n(a_{*})(a-a_{*})^n=\Omega_{M0}^{\frac 1 2}H_0(t-t_{*}).
\end{equation}
The coefficients of the expansion are
\begin{eqnarray}
\tilde{c}_1(a_{*})&=&-\frac{a_{*}^{\frac 1 2}\Omega_{M0}^{\frac 12 }}{\nu},\nonumber\\
\tilde{c}_n(a_{*})&=&-\frac{\tilde{A}_n(a_{*})c_1(a_{*})}{2^{n-1}(n!)\nu^{2(n-1)}a_{*}^{n-1}},\quad (n\geq 2)
\end{eqnarray}
where $\nu =\sqrt{\Omega_{M0}+\Omega_{\Lambda 0}a_{*}^3}, \Omega_{\Lambda 0}=1-\Omega_{M0}$ and
\begin{eqnarray}
\tilde{A}_2(a_{*})&=&2\Omega_{M0}-4\Omega_{\Lambda 0}a_{*}^3,\nonumber\\
\tilde{A}_3(a_{*})&=&\Omega_{M0}^2+20\Omega_{\Lambda 0}\Omega_{M0}a_{*}^3-8\Omega_{\Lambda 0}^2a_{*}^6,\nonumber\\
\tilde{A}_4(a_{*})&=&3(2\Omega_{M0}^3-28\Omega_{\Lambda 0}\Omega_{M0}^2a_{*}^3+208\Omega_{\Lambda 0}^2\Omega_{M0}a_{*}^6\nonumber\\
&-&32\Omega_{\Lambda 0}^3a_{*}^9),\nonumber\\
\tilde{A}_5(a_{*})&=&3(5\Omega_{M0}^4+40\Omega_{\Lambda 0}\Omega_{M0}^3a_{*}^3-1008\Omega_{\Lambda 0}^2\Omega_{M0}^2a_{*}^6\nonumber\\
&+&1664\Omega_{\Lambda 0}^3\Omega_{M0}a_{*}^9-128\Omega_{\Lambda 0}^4a_{*}^{12}),\cdots\cdots
\end{eqnarray}

Note that $t_{*}$ in Eqs.(60) and (63) can be determined by use of any known evolving time in the convergent range of the series. For example, $t_{*}$ can be expressed as
\begin{equation}
t_{*}=\sum_{n=1}^{\infty}\frac{(-1)^nc_n(\frac 1 7)(a_{*}-\frac 1 7)^n}{\Omega_{M0}^{\frac 1 2}H_0}+t_{z=6}.
\end{equation}

Weierstrass\cite{Ahlfors} had built the whole theory of analytic functions from the concept of power series. The process above can be repeated any times, each of which is a direct analytic continuation of the preceding one. By using this method we can describe analytically the whole evolution history of the Universe.

We define $a(t_{crit})=a_{crit}$ is the critical values of the scale factor of the two models through which the Universe changes to the acceleration phase from deceleration one.
From Eqs. (61)-(63), we find that $a_{crit}$ should satisfy the equation $c_2(a_{crit})=0$ so that
\begin{widetext}
\begin{equation}
a_{crit}=\begin{cases}\Big ( \frac{1} {6B}(A^2+15\Omega_{M0}^2)^{\frac 1 2}\cos \left (\frac 1 3\cos^{-1}\frac{A^3+63AB\Omega_{M0}^2}{(A^2+15B\Omega_{M0}^2)^{3/2}}\right )-\frac A {6B}\Big )^{\frac 1 3}\\\left (\frac{2\Omega_{M0}^2}{A}\right )^{\frac 1 3}\end{cases}\quad \text{for} \begin{array}{c}\text{Model I}\\\text{Model II}\end{array}.
\end{equation}
\end{widetext}
Certainly, $a_{crit}$ can be straightforwardly obtained from the expression of deceleration parameter (see Eq. (70)). Using fitting values, we have
\begin{equation}
a_{crit}=\begin{cases}0.599\\0.752\end{cases}\quad \text{for} \begin{array}{c}\text{Model I}\\\text{Model II}\end{array}.
\end{equation}
Similarly, we have $a_{crit}^{(\Lambda)}=0.603$ for $\Lambda$CDM. Substituting Eqs. (61) and (68) into Eq. (66), we have the critical time as
\begin{equation}
t_{crit}=\begin{cases}0.526H_0^{-1}\\0.616H_0^{-1}\end{cases}\quad \text{for} \begin{array}{c}\text{Model I}\\\text{Model II}\end{array}.
\end{equation}
Furthermore, the unifiable form of the deceleration parameter of the two models can be written as
\begin{equation}
q=\frac 1 2-\frac{3Aa^3+6Ba^6}{\Omega_{M0}^2+4Aa^3+4Ba^6+\Omega_{M0}\sqrt{\Omega_{M0}^2+4Aa^3+4Ba^6}}.
\end{equation}
The evolutions of $q$ for the two models are plotted in Fig.4.
\begin{figure}[h]
\begin{center}
\includegraphics[width=0.4\textwidth,angle=0]{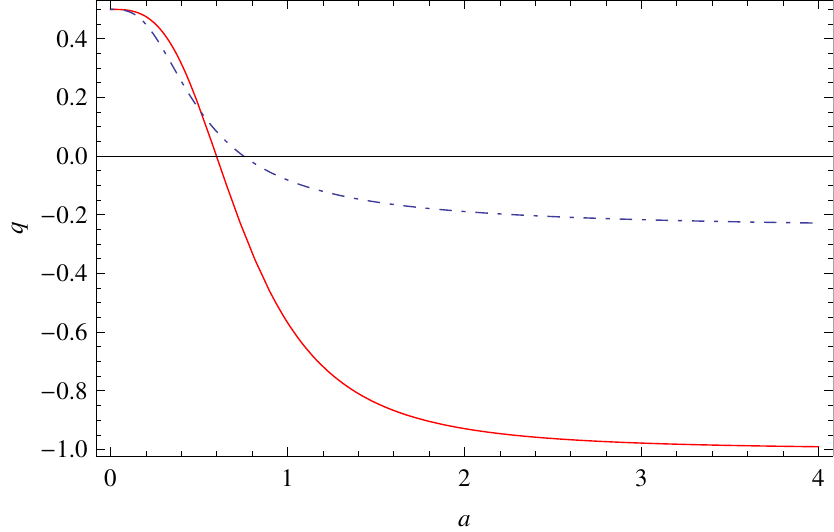}
\caption{The evolution of the deceleration parameter. The solid line is for model I and the dotdashed is for model II.}
\end{center}
\end{figure}

The age of the Universe can be rewritten as
\begin{equation}
t_0=\sum_{n=1}^{\infty}\frac{c_n(a_{crit})(1-a_{crit})^n}{\Omega_{M0}^{\frac 1 2}H_0}+t_{crit}.
\end{equation}
Using the fitting values, we have
\begin{equation}
t_0=\begin{cases}0.966H_0^{-1}\approx13.77 \text{Gyr}\\0.866H_0^{-1}\approx13.88 \text{Gyr}\end{cases}\quad \text{for} \begin{array}{c}\text{Model I}\\\text{Model II}\end{array}.
\end{equation}

According to Ref.\cite{Marin-Franch}, the ages of the globular clusters are
\begin{equation}
t_0=12.8\pm 0.4 \text{Gyr}.
\end{equation}
Therefore, the values of Eq. (72) are consistent with the ages of the oldest globular clusters.

From Eqs. (60)-(65), we have the difference of scale factor $\Delta a$ at time $t$ between our models and $\Lambda $CDM,
\begin{widetext}
\begin{equation}
\sum_{n=1}^{\infty}(-1)^n\{\Delta_ic_n(a_{crit}^{(\text{I,II})})\left [a(1+\Delta a)-a_{crit}^{(\text{I,II})}\right ]^n-c_n(a_{crit}^{(\Lambda)})(a-a_{crit})^n\}=(\Omega_{M0}^{\frac 1 2}H_0)^{(\Lambda)}(t_{crit}^{(\Lambda)}-\frac 1 {\Delta_i} t_{crit}^{(\text{I,II})}).
\end{equation}
\end{widetext}
Using Eqs. (69) and (74), and the fitting values, we have
\begin{equation}
7.38\times 10^{-4}<\Delta a<1.62\times 10^{-3} \quad \text{for model I},
\end{equation}
and
\begin{equation}
-3.33\times 10^{-2}>\Delta a>-1.06\times 10^{-2} \quad \text{for model II},
\end{equation}
In the redshift interval $0<z<6$. Therefore, our models do not change the $\Lambda$CDM scenario under the error band of fitting values for the observed data.

\subsection{Fate of the Universe}
In the future of the Universe, the evolutions of the scale factor are calculated and illustrated in Fig.5.
\begin{figure}[h]
\begin{center}
\includegraphics[width=0.4\textwidth,angle=0]{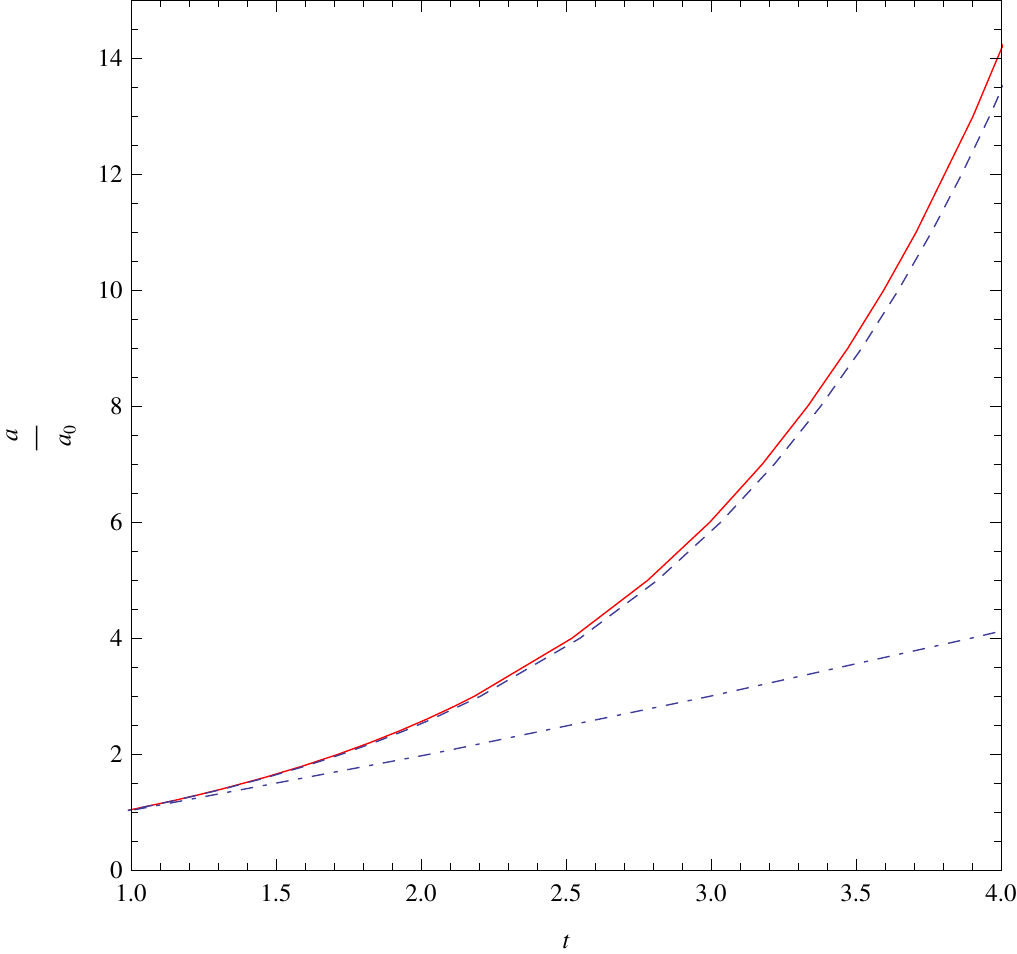}
\caption{ The evolutions of the scale factor in the future of the Universe. The dotted line is for $\Lambda$CDM, the solid line is for model I and the dotdashed line is for model II.}
\end{center}
\end{figure}

In the far future, when $ t\gg t_0$ and $a\gg 1$, we have

\begin{equation}
q=\begin{cases}-1\\-\frac 1 4 \end{cases}\quad \text{for} \begin{array}{c}\text{Model I}\\\text{Model II}\end{array},
\end{equation}

\begin{equation}
w_{DE}=\begin{cases}-1\\-\frac 1 2 \end{cases}\quad \text{for} \begin{array}{c}\text{Model I}\\\text{Model II}\end{array},
\end{equation}

\begin{equation}
a(t)=\begin{cases}\exp [B^{\frac 1 4}H_0(t-t_0)]\\\left(\frac 3 4H_0A^{\frac 1 4}(t-t_0)\right )^{\frac 4 3} \end{cases}\quad \text{for} \begin{array}{c}\text{Model I}\\\text{Model II}\end{array}
\end{equation}
and
\begin{equation}
\ddot{a}(t)=\begin{cases}B^{\frac 1 2}H_0^2\exp [B^{\frac 1 4}H_0(t-t_0)]\\(6^{-2}H_0^4A)^{\frac 1 3}(t-t_0)^{-\frac 2 3} \end{cases}\quad \text{for} \begin{array}{c}\text{Model I}\\\text{Model II}\end{array}.
\end{equation}
Therefore, the acceleration of expansion tends to zero (infinite) as $t\rightarrow \infty $ for model II ( model I). It is apropos that the deceleration parameter $q$ will be ineffective when $a(t)$ tends to infinite or zero.  Obviously, the fates of the Universe are de Sitter expansion and power law expansion for model I and II, respectively, and there are no future singularities in both of our models, even though $w_{DE0}=-1.019\pm 0.035$ makes model I look like a phantom at present time.

 \section{conclusion and discussion}

The clear advantage of $f(T)$ theory that the field equations are second order but not fourth order as in $f(R)$ theory can afford us to make things convenient for establishing cosmological models. The extension of $f(T)$ gravity with a nonminimal torsion-matter coupling has rich cosmological implications. In this paper, we have studied two concrete $f(T)$ models with nonminimal torsion-matter coupling extension using several observation data including SNeIa, CMB and BAO. When describing the evolution history of the Universe including radiation-dominated era, matter-dominated era and the present accelerating expansion, our models are in consistent with the successful aspects of $\Lambda$CDM scenario under the error band of fitting values, while with the superiority that they alleviate the cosmological constant problem. The best fitting results $A=0.188$, $B=0.510$ for model I and $A=0.633$ for model II but not $H_0$ determine the present values of $f_1$ and $f_2$, which indicate the lucky avoidance of our models from fine tuning. Furthermore, the age of the Universe obtained from our models is consistent with the ages of the oldest globular clusters.

Despite the $f(T)$ models I and II being highly successful in describing the observations of the Universe, its large scale structure and evolution, they may have same problems in describing structures at small scales as $\Lambda$CDM scenario \cite{Moore,Ostriker,Boylan}. The solutions to the structure problems at small scales including the cusp/core problem and the missing satellite problem (MSP) can be distinguished into cosmological and astrophysical solutions. Cosmological solutions can be based on modifying the power spectrum at small scales \cite{Zentner}. Ferraro has made an attempt on modification of the spectrum of perturbations in a $f(T)$ theory \cite{Ferraro2}. An interesting question is how the spectrum of the perturbation at small scales will be modified in our models. This will be considered in our next work.

On the other hand, since GR is in excellent agreement with Solar system and binary pulsar observations\cite{Will}, any modified gravity theory aiming at explaining the large-scale dynamics of the Universe have to reproduce GR in weak-field limit to ensure its viability. Recently, Solar system data\cite{Iorio,Xie} have been used to constrain $f(T)$ theories with a power-law ansatz for an additive term to the TEGR Lagrangian. But it is noticed that the choice of a good tetrad is crucial to meet the Lorentz invariance, and a nondiagonal tetrad is good for spherical coordinates, while a diagonal one is good for Cartesian coordinates\cite{Tamanini}. Later on, the weak-field spherically symmetric solutions in $f(T)$ gravity is reexamined with the consideration of choosing good tetrad and give constraints to $f(T)$ gravity\cite{Ruggiero} . We should also investigate the Solar system tests and give constraints for our models. This issue will be addressed in a separated paper.

Another interesting question is if there are other concrete models satisfying observation data. Certainly there are. However, we find no acceptable $H_0$ and $\Omega_{m0}$ when $f_1 (T)$ and $f_2 (T)$ are both proportional to $T^2$. Furthermore, if we restrict ourselves to consider power law models $f_1(T)\propto T^{\alpha}$ and $f_2(T)\propto T^{\beta}$, the range of $\alpha$ and $\beta$ may be given by use of the observation data of SNeIa, BAO and CMB. We will consider this question in the future work.

\begin{acknowledgments}
This work is supported by National Science Foundation of China grant Nos.~11105091 and~11047138, ``Chen Guang" project supported by Shanghai Municipal Education Commission and Shanghai Education Development Foundation Grant No. 12CG51, National Education Foundation of China grant  No.~2009312711004, Shanghai Natural Science Foundation, China grant No.~10ZR1422000, Key Project of Chinese Ministry of Education grant, No.~211059,  and  Shanghai Special Education Foundation, No.~ssd10004, and the Program of Shanghai Normal University.
\end{acknowledgments}

\end{document}